\documentclass[twocolumn,pra,aps,superscriptaddress]{revtex4}
\usepackage[T1]{fontenc}
\usepackage[latin9]{inputenc}
\usepackage{amsmath}
\usepackage{amssymb}
\usepackage{graphicx}
\usepackage{esint}

\usepackage[english]{babel}
\usepackage{times}
\usepackage{booktabs}
\usepackage{array}
\usepackage{supertabular}
\usepackage{multirow}
\usepackage{calc}
\usepackage{mathtools}
\usepackage{bm}
\usepackage{stackrel}
\usepackage{setspace}
\usepackage{subfigure}
\usepackage{indentfirst}
\usepackage{cases}
\usepackage[colorlinks,linkcolor=blue,citecolor=blue,hyperindex,bookmarks=false,pdfstartview=FitH,dvipdfm]{hyperref}



\begin{document}
\title{Spatial enantioseparation of gaseous chiral molecules}

\author{Bo Liu}
\affiliation{Beijing Computational Science Research Center, Beijing 100193, China}

\author{Chong Ye}
\email{yechong@bit.edu.cn}
\affiliation{Beijing Key Laboratory of Nanophotonics and Ultrafine Optoelectronic Systems School of Physics, Beijing Institute of Technology, Beijing 100081, China}
\affiliation{Beijing Computational Science Research Center, Beijing 100193, China}

\author{C. P. Sun}
\affiliation{Beijing Computational Science Research Center, Beijing 100193, China}
\affiliation{Graduate School of China Academy of Engineering Physics, No. 10 Xibeiwang East Road, Haidian District, Beijing 100193, China}

\author{Yong Li}
\email{liyong@csrc.ac.cn}
\affiliation{Beijing Computational Science Research Center, Beijing 100193, China}
\affiliation{Synergetic Innovation Center for Quantum Effects and Applications, Hunan Normal University, Changsha 410081, China}

\begin{abstract}
We explore the spatial enantioseparation of gaseous chiral molecules for the cyclic three-level systems coupled with three electromagnetic fields. Due to molecular rotations, the specific requirements of the polarization directions of the three electromagnetic fields lead to the space-dependent part of the overall phase of the coupling strengths. Thus, the overall phase of the coupling strengths, which differs with $\pi$ for the enantiomers in the cyclic three-level model of chiral molecules, varies intensely in the length scale of the typical wavelength of the applied electromagnetic fields. Under the induced gauge potentials resulting from the space-dependent part of the overall phase and the space-dependent intensities of coupling strengths, we further show spatial enantioseparation for typical parameters of gaseous chiral molecules.
\end{abstract}

\maketitle

\section{Introduction} \label{Introduction}
The left- and right-handed chiral molecules (called enantiomers) coexist in many biologically active compounds. Usually, only one form of enantiomer is biologically beneficial, while the other one is useless or even harmful.
Thus, the enantioseparation of chiral molecules is fundamentally
significant in organic chemistry~\cite{Latscha}, pharmacology~\cite{Hutt,Ariens,Eriksson,Teo}, and biochemistry~\cite{Gal,Leitereg,Hyttel}.
It also becomes an important topic in atomic, molecular, and optical physics~\cite{Zhang,Zhang2,Zhang3,Barcellona,Barcellona2,Gershnabel,Tutunnikov,Forbes,Forbes2,Forbes3,Yachmenev,Yurchenko,Cameron,Bradshaw,Bradshaw2,Shapiro,Salam,Fujimuraa,Shapiro3,Shapiro4,Drewsen,Wu2}.
Among these methods~\cite{Zhang,Zhang2,Zhang3,Barcellona,Barcellona2,Gershnabel,Tutunnikov,Forbes,Forbes2,Forbes3,Yachmenev,Yurchenko,Cameron,Bradshaw,Bradshaw2,Shapiro,Salam,Fujimuraa,Shapiro3,Shapiro4,Drewsen,Wu2}, an interesting one is
based on a cyclic three-level ($\Delta$-type) configuration~\cite{Shapiro,Salam,Fujimuraa,Shapiro3,Shapiro4,Drewsen,Wu2},
which is formed by applying three electromagnetic fields to couple with the electric dipole transitions among three levels.
In general, such a cyclic three-level system is forbidden in natural atoms
whose states have certain (even or odd) parities,
but it exists in chiral molecules and other symmetry-broken systems~\cite{Ansari,Blockley,Liu,WangZH2,Zhou}.
For chiral molecules, the cyclic three-level model is special since the product of the three corresponding coupling strengths of the
electric dipole transition moments changes sign with enantiomers.
Then the overall phase of the product of the three corresponding coupling strengths differs by $\pi$ with two enantiomers~\cite{Kral2,Kral}.
This property makes the cyclic three-level model feasible in the enantioseparation~\cite{Kral,Li3,Jia2,Ye1,Li2,ShapiroLiX,Torosov1,Torosov2,Wu} and the enantiodiscrimination~\cite{Hirota,Lehmann4,Jia1,Ye2,Chen,Xu,Kang,Chen2} of chiral molecules.

In the cyclic three-level systems, the inner-state enantioseparation (commonly called  enantio-specific state transfer)~\cite{Kral,Li3,Jia2,Ye1,Torosov1,Torosov2,Wu} and the spatial enantioseparation~\cite{Li2,ShapiroLiX} have been discussed theoretically. The corresponding experiments of inner-state enantioseparation have also been investigated in experiments~\cite{Eibenberger,Perez}.
For inner-state enantioseparation, a chiral mixture can be enantiopurified in one of the three inner states (i.e., with only one enantiomer occupying that state)~\cite{Kral,Li3,Jia2}.
The enantiopure molecules in that state can be further spatially separated from the initial chiral mixture by a variety of energy-dependent processes.
For spatial enantioseparation~\cite{Li2,ShapiroLiX},
the different overall phases of the coupling strengths for two enantiomers lead to the chirality-dependent induced gauge potentials for molecules.
By these induced gauge potentials, a racemic molecular beam is split into subbeams, depending on the molecular chirality
or inner state.

For real molecules, especially the gaseous chiral molecules used in recent experiments~\cite{Eibenberger} for the cyclic three-level systems, the molecular rotations should be considered and may introduce some negative effects on inner-state or spatial enantioseparation in the original schemes~\cite{Kral2,Kral,Li3,Jia2,Ye1,Li2,ShapiroLiX,Torosov1,Torosov2,Wu}.
Due to the magnetic degeneracy of the molecular rotational states, the ideal single-loop cyclic three-level system~\cite{Kral2,Kral,Li3,Jia2,Ye1,Li2,ShapiroLiX,Torosov1,Torosov2,Wu} is generally replaced by a complicated multiple-loop one~\cite{Hornberger}.
This reduces the ability of inner-state enantioseparation and spatial enantioseparation of chiral molecules.
Fortunately, for the asymmetric-top chiral molecules, the single-loop configuration is constructed
by applying three electromagnetic fields with appropriate polarization vectors and frequencies~\cite{Ye,Koch}.

In this paper, for the single-loop cyclic three-level system, we discuss the spatial enantioseparation of asymmetric-top chiral molecules taking into consideration molecular rotations.
In order to construct the single-loop cyclic configuration, the propagation directions of three electromagnetic fields
may not be parallel. Then, under the three-photon resonance condition, molecules in different spatial positions ($\vec{r}$) experience different space-dependent parts ($\phi_{r}=\delta\vec{k} \cdot \vec{r}$~\cite{dk}) of the overall phases of the coupling strengths, which results in the problem of phase mismatching~\cite{Lehmann} in the inner-state enantioseparation~\cite{Kral,Li3,Jia2,Ye1,Torosov1,Torosov2,Wu,Eibenberger,Perez} and the enantiodiscrimination~\cite{Patterson,Patterson2,Shubert,Shubert2,Shubert3,Lobsiger,Patterson3}. We find that this space-dependent part of the overall phase (besides the spatial distribution of the intensity of the three coupling strengths) offers an additional resource that results in induced gauge potentials.
In contrast, in the original
research of the spatial enantioseparation
of cyclic three-level systems~\cite{Li2,ShapiroLiX} (as well as other research of the inner-state enantioseparation~\cite{Kral,Li3,Jia2,Ye1,Torosov1,Torosov2,Wu}), the molecular rotations were not considered.
Then the propagation directions of three electromagnetic fields are designed to be parallel for simplicity. Under the three-photon resonance condition, the space-dependent part $\phi_{r}$ of the overall phase of the coupling strengths is zero. So, in this case~\cite{Li2,ShapiroLiX}, the induced gauge potentials originate only from the spatial distribution of the intensity of three coupling strengths.

Specifically, we pay attention to the induced gauge potentials in two special directions, i.e.,
in the $\delta\vec{k}$ direction (called the $\hat{q}$ direction) and perpendicular to the $\delta\vec{k}$ direction (called the $\hat{p}$ direction).
We make the transverse profile of each electromagnetic field have the Gaussian form.
The radii of the Gaussian beams are chosen to be much larger than the wavelength of the electromagnetic field.
Under the above condition,
the induced gauge potentials in the $\hat{q}$ direction
exhibit the slowly varying chirality-independent envelope multiplied by a quickly varying chirality-dependent quasi periodic function.
The induced gauge potentials in the $\hat{p}$ direction are chirality dependent and slowly varying, similar to the previous case without considering the molecular rotations~\cite{Li2}.

Finally,
we explore the center-of-mass motion of chiral molecules under such induced gauge potentials for two cases with initial velocity in the $\hat{q}$ and $\hat{p}$ directions, respectively.
In a short time so that the molecules move only  the distance of a typical wavelength scale, we find
that the effect of the spatial enantioseparation in the former case is much better than the effect in the latter case.
In a long time  in which the molecules move a distance much longer than a typical wavelength, the effect of the spatial enantioseparation is comparable in the two cases that the initial velocity in the $\hat{q}$ and $\hat{p}$ directions.

\section{Cyclic three-level model with the space-dependent part of the overall phase} \label{Model}
\begin{figure}[htbp]
\centering
\includegraphics[width=0.85\linewidth]{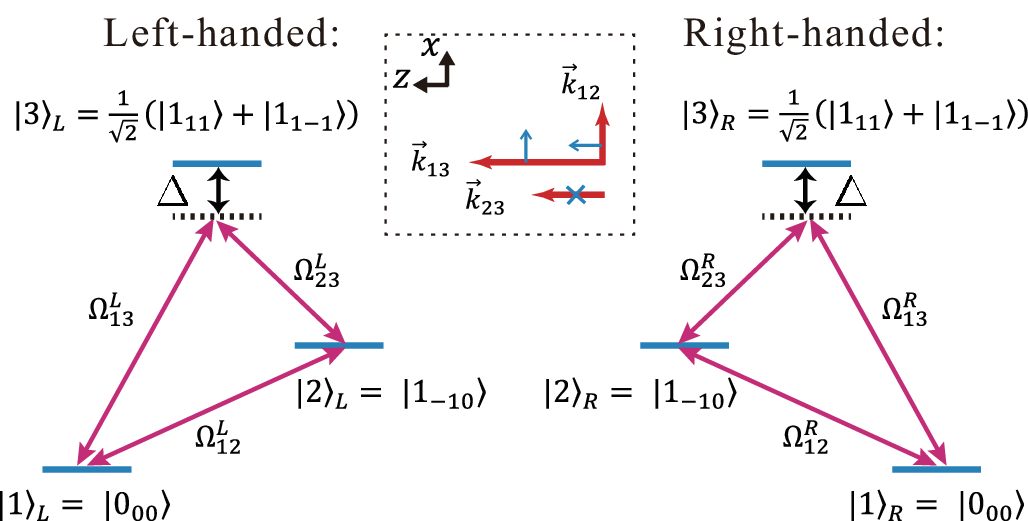}
\caption{
Model of cyclic three-level ($\Delta$-type) left-handed and right-handed chiral molecules, coupled to the electromagnetic fields with
the intensities $\Omega_{jl}^{L,R}$ of the coupling strengths.
The inset shows the directions of propagation (in red) and polarization (in blue) for the three electromagnetic waves.
\label{fig:model}}
\end{figure}

We consider the cyclic three-level ($\Delta$-type) model composed of three inner states ($\left|1\right\rangle$, $\left|2\right\rangle$, and $\left|3\right\rangle$) of chiral molecules as shown in Fig.~\ref{fig:model}.
The wave function of a single molecule takes the form
\begin{equation}
\left|\Psi\left(\vec{r}\right)\right\rangle =\sum_{j=1}^{3}\psi_{j}\left(\vec{r}\right)\left|j\right\rangle
\end{equation}
and the Hamiltonian is
\begin{equation}
H=\frac{1}{2m}\left(-\mathrm{i} \hbar\nabla \right)^{2}+H_{\text{inn}},
\end{equation}
where $\left(-\mathrm{i} \hbar\nabla\right)^{2}/2m$ is the kinetic energy of the center-of-mass motion and the inner Hamiltonian $H_{\text{inn}}$ reads
\begin{eqnarray}
H_{\text{inn}}
& = & \sum_{j=1}^{3} \hbar\omega_{j}\left|j\right\rangle \left\langle j\right|  \label{eq:Hinn} \\
&   & +\sum_{l>j=1}^{3}\hbar \left[ \Omega_{jl}\exp\left(\mathrm{i}\omega_{jl}t-\mathrm{i}\vec{k}_{jl}\cdot\vec{r}\right)\left|j\right\rangle \left\langle l\right|+\text{H.c.}\right]. \nonumber
\end{eqnarray}
Here $\hbar \omega_{j}$ correspond to the inner level energies,
$\vec{k}_{jl}$ are the wave vectors,
$\vec{r}$ is the position of the molecular center of mass,
$\omega_{jl}$ are the frequencies of the electromagnetic fields, and
$\Omega_{jl}\exp(\mathrm{i}\omega_{jl}t-\mathrm{i}\vec{k}_{jl}\cdot\vec{r})$ are the coupling strengths with
$\Omega_{jl}=\vec{\mu}_{jl}\cdot\vec{E}_{jl}$.
Here $\vec{\mu}_{jl}$ are the electrical dipole matrix elements and $\vec{E}_{jl}$ are the electromagnetic fields coupling with the transition $\left|j\right\rangle \leftrightarrow \left|l\right\rangle$.

The chirality dependence of the molecule is reflected as~\cite{Kral,Kral2}
\begin{equation}
\Omega_{12}^{L}=-\Omega_{12}^{R},~~\Omega_{13}^{L}=\Omega_{13}^{R},~~\Omega_{23}^{L}=\Omega_{23}^{R}.
\label{eq:Omega}
\end{equation}
Here we have added the superscript $L$  or $R$ to denote the left-handed or right-handed chiral molecule.
When referring to left-handed or right-handed chiral molecules, we add the superscript.
When there is no superscript, we are referring to general molecules.
It is clear from Eq.~(\ref{eq:Omega}) that the overall phase of the product of the three coupling strengths
for two enantiomers at the same position has a difference of $\pi$~\cite{Kral}.
Such a chirality dependence offers the possibility to realize the enantioseparation~\cite{Kral,Li3,Jia2,Ye1,Li2,ShapiroLiX,Torosov1,Torosov2,Wu,Hornberger,Eibenberger,Ye,Koch,Perez}, the
enantiodiscrimination~\cite{Hirota,Patterson3,Lehmann4,Jia1,Ye2,Chen,Xu,Kang,Chen2,Patterson,Patterson2,Shubert,Shubert2,Shubert3,Lobsiger}, and the
enantioconversion~\cite{Brumer,Gerbasi,Frishman,Ye3,Ye4}.
Besides the chirality-dependent part of the overall phase, the overall phase usually contains the space-dependent part $\phi_{r}$ of the overall phase and the time-dependent part $\phi_{t}$ of the overall phase:
\begin{eqnarray}
\phi_{r} & = & \delta\vec{k} \cdot \vec{r} \equiv \left(\vec{k}_{13}-\vec{k}_{23}-\vec{k}_{12} \right) \cdot \vec{r},\\
\phi_{t} & = & \left(\omega_{12}+\omega_{23}-\omega_{13}\right)t.\nonumber
\end{eqnarray}
Usually, we take the three-photon resonance condition
\begin{equation}
\omega_{12}+\omega_{23}-\omega_{13} = 0.
\label{eq:nu}
\end{equation}
Then, the time-dependent part $\phi_{t}$ of the overall phase is zero.

In the original investigations of the (inner-state or spatial) enantioseparation~\cite{Kral,Li3,Jia2,Ye1,Li2,ShapiroLiX,Torosov1,Torosov2,Wu}, the molecular rotations are not considered, so the wave vectors of the three electromagnetic fields ($\vec{k}_{12}$, $\vec{k}_{23}$, and $\vec{k}_{13}$) are parallel to each other.
Then, the phase matching is satisfied (i.e., $\delta\vec{k}=\vec{k}_{13}-\vec{k}_{23}-\vec{k}_{12}=0$), and
the corresponding space-dependent part $\phi_{r}$ of the overall phase is zero.
However, for gaseous chiral molecules, the molecular rotations should be considered.

Now we consider the (single-loop) cyclic three-level models of gaseous chiral molecules by considering the rotational degrees of freedom, following the methods in Refs.~\cite{Hornberger,Eibenberger}.
We consider different rotational states $\left|J_{\tau M}\right\rangle$ with the same vibrational ground state $\left|v_{g}\right\rangle$. Here
$J$ is the angular momentum quantum number, $M$ is the magnetic quantum number,
and $\tau$ runs from $-J$ to $J$ in unit steps in the order of increasing energy~\cite{Zare}.
We omit $\left|v_{g}\right\rangle$ in further discussions.
We choose the working states as $\left|1\right\rangle= \left|0_{00}\right\rangle$,
$\left|2\right\rangle= \left|1_{-10}\right\rangle$,
and $\left|3\right\rangle=(\left|1_{11}\right\rangle+\left|1_{1-1}\right\rangle)/\sqrt{2}$ (see Fig.~\ref{fig:model}).
Correspondingly, these three electromagnetic fields are $\hat{z}-$, $\hat{x}-$ and $\hat{y}-$polarized.
Then, the wave vectors ($\vec{k}_{12}$, $\vec{k}_{23}$, and $\vec{k}_{13}$) of the three electromagnetic fields may not be parallel.
For simplicity, we make $\vec{k}_{12}=|\vec{k}_{12}|\,\hat{x}$ propagate in the $\hat{x}$ direction and
$\vec{k}_{23}=|\vec{k}_{23}|\,\hat{z}$ and $\vec{k}_{13}=|\vec{k}_{13}|\,\hat{z}$ propagate in the $\hat{z}$ direction (see the inset of Fig.~\ref{fig:model}).

Such a setup of the electromagnetic fields makes $\delta\vec{k}\ne 0$,
which means that the space-dependent part $\phi_{r}$ of the overall phase is nonzero.
This introduces the problem of phase mismatching, which is the main limitation of the inner-state enantioseparation~\cite{Perez} and the
enantiodiscrimination~\cite{Patterson,Patterson2,Shubert,Shubert2,Shubert3,Lobsiger,Patterson3}.
In the following we discuss how the nonzero space-dependent part $\phi_{r}$ of the overall phase affects the spatial enantioseparation of gaseous chiral molecules.

\section{Induced gauge potentials: quasi-periodic structure} \label{Induced VA}
In the preceding section we showed  our cyclic three-level model taking into consideration molecular rotations.
In this section, we show the induced gauge potentials of such a model.

To this end, we assume the conditions of one-photon resonance
$\left(\omega_{2}-\omega_{1}\right)-\omega_{12} = 0$
and large detuning as well as weak coupling
$\left|\Delta\right|\gg\left|\Omega_{13}\right|\sim\left|\Omega_{23}\right|\gg\left|\Omega_{12}\right|$
with $\Delta\equiv\left(\omega_{3}-\omega_{1}\right)-\omega_{13}=\left(\omega_{3}-\omega_{2}\right)-\omega_{23}$.
By adiabatically eliminating the excited state $\left|3\right\rangle$ in the large detuning condition~\cite{Li2},
we have the effective inner Hamiltonian
\begin{equation}
H_{\text{inn}}^{\prime} =
\hbar \Lambda_{1} \left|1\right\rangle \left\langle 1\right|
+\hbar \Lambda_{2} \left|2\right\rangle \left\langle 2\right|
+\hbar \left(g e^{\mathrm{i} \Phi}\left|1\right\rangle \left\langle 2\right|+\text{H.c.}\right)
\label{eq:Heffinn}
\end{equation}
with
\begin{equation}
g e^{\mathrm{i} \Phi}\equiv e^{-\mathrm{i}\left|\vec{k}_{12}\right|z}\left(\Omega_{12}
e^{\mathrm{i}\phi_{r}}
-\frac{\Omega_{13}\Omega_{23}^{\ast}}{\Delta}\right). \label{eq:gPhi}
\end{equation}
Here $g$ is a positive real constant and $\Phi$ is a real constant.
They are both uniquely determined by the parameters on the
right-hand side of Eq.~(\ref{eq:gPhi}). Actually, they will depend on the chirality of the (left- or right-handed) chiral molecules
(with $\Omega_{jl} \rightarrow \Omega_{jl}^{L,R}$).
Then we have the inner dressed states~\cite{Li2}
\begin{eqnarray}
|\chi_{1}\rangle
& = & \cos\theta\left|1\right\rangle +e^{-\mathrm{i} \Phi}\sin\theta\left|2\right\rangle, \nonumber \\
|\chi_{2}\rangle
& = & -\sin\theta\left|1\right\rangle +e^{-\mathrm{i} \Phi}\cos\theta\left|2\right\rangle, \label{eq:Eigvalues}
\end{eqnarray}
with the corresponding eigenvalues
$\varepsilon_{1}=\hbar (\Lambda_{1}+g\tan\theta)$ and
$\varepsilon_{2}=\hbar (\Lambda_{2}-g\tan\theta)$,
respectively. Here
$\Lambda_{1} = -|\Omega_{13}|^{2}/\Delta$,
$\Lambda_{2} = -|\Omega_{23}|^{2}/\Delta$, and $\theta$ is given by $\tan2\theta=2g/(\Lambda_{1}-\Lambda_{2})$.
The parameters $\theta$ and $\Phi$  are related to
the intensities and space-dependent part $\phi_{r}$ of the overall phase of the coupling strengths.
Meanwhile, they are both chirality dependent and space dependent.

In the inner dressed states basis
$\{ |\chi_{1}\rangle,\,|\chi_{2}\rangle \}$,
we assume the adiabatic approximation~\cite{Li2,ShapiroLiX}.
Then the two-level system reduces to two subsystems
$H^{\prime} = \sum_{\alpha=1}^{2} H_{\alpha}|\chi_{\alpha}\rangle\langle \chi_{\alpha}|$
with~\cite{C.P.Sun,Zhu,Dalibard,Ruseckas}
\begin{equation}
H_{\alpha} = \frac{1}{2m}(-\mathrm{i} \hbar\nabla-\vec{A}_{\alpha} )^{2}+V_{\alpha}.
\label{eq:Halpha}
\end{equation}
Their corresponding induced vector potentials $\vec{A}_{\alpha}$ and the induced scalar potentials $V_{\alpha}$ are
\begin{eqnarray}
\vec{A}_{\alpha}
& = & \mathrm{i} \hbar \left\langle \chi_{\alpha} \left|\nabla \chi_{\alpha} \right.\right\rangle, \nonumber\\
V_{\alpha}
& = & \varepsilon_{\alpha},~~(\alpha=1,2).
\label{eq:A and V}
\end{eqnarray}
For real $\Omega_{12}$, $\Omega_{23}$, and $\Omega_{13}$, the induced scalar potentials are
$V_{1}=-R+W$ and $V_{2}=-R-W$, with $R=(\Omega_{13}^{2}+\Omega_{23}^{2})/2\Delta$ and $W=(\Omega_{12}^{2}+R^{2}
-2\cos\phi_{r}\Omega_{12}\Omega_{23}\Omega_{13}/\Delta)^{1/2}$. It is clear that the induced vector and scalar potentials depend on both the chirality and inner dressed states~\cite{Li2,ShapiroLiX} and change with the spatial position.
We note that the subsequent results based on the Hamiltonian~(\ref{eq:Halpha}) with U(1) gauge fields are gauge invariant~\cite{C.P.Sun,sun}.

It is observed from Eqs.~(\ref{eq:gPhi}-\ref{eq:A and V}) that
the nonzero space-dependent part $\phi_{r}$ of the overall phase also contributes to the induced gauge potentials.
This is very different from the original research~\cite{Li2,ShapiroLiX} without considering the molecular rotations, where the induced gauge potentials are only attributed to the spatial distribution of the intensities of the three coupling strengths.

In order to show how the space-dependent part $\phi_{r}$ of the overall phase affects the induced gauge potentials, we choose a new coordinate with
$\hat{q}=(\hat{z}-\hat{x})/\sqrt{2}$ and $\hat{p}=(\hat{z}+\hat{x})/\sqrt{2}$.
Since $\delta\vec{k}=|\vec{k}_{12}|(\hat{z}-\hat{x})=\sqrt{2}|\vec{k}_{12}|\hat{q}$ is in the $\hat{q}$ direction,
the space-dependent part $\phi_{r}$ of the overall phase only varies in the $\hat{q}$ direction.

We take the 1,2-propanediol as an example as follows.
The corresponding bare transition frequencies of the related three-level model are given by
$\omega_{21} /2\pi = 6431.06  \, \text{MHz}$,
$\omega_{32} /2\pi = 5781.09  \, \text{MHz}$, and
$\omega_{31} /2\pi = 12212.15 \, \text{MHz}$~\cite{Lovas}, respectively,
where $\omega_{jl}=|\omega_{j}-\omega_{l}|$.
The intensities $\Omega_{jl}$ of the coupling strengths are assumed to have the Gaussian forms
\begin{figure}[htbp]
\centering
\includegraphics[width=0.9\columnwidth]{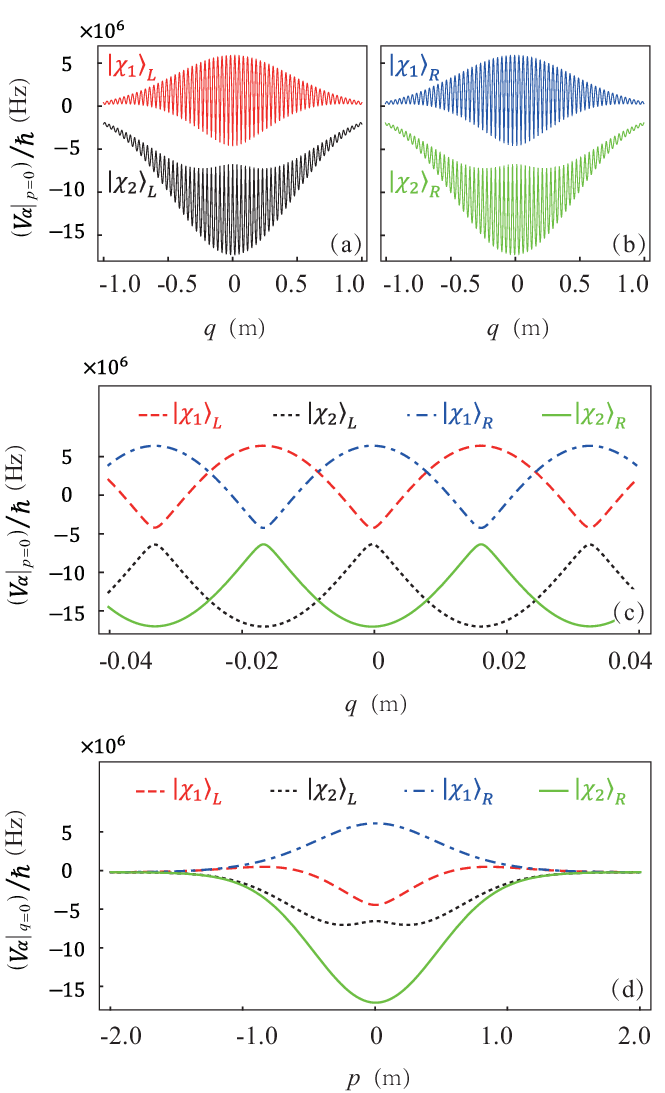}
\caption{
Induced scalar potentials $V_{\alpha}$ at fixed $p=0$ (expressed as $V_{\alpha}|_{p=0}$)
in the $\hat{q}$ direction for (a) a left-handed molecule and (b) a right-handed molecule.
 (c) Regions $q\in [-0.04,\,0.04]\,\text{m}$ of (a) and  (b) are enlarged and shown together.
(d) Induced scalar potentials $V_{\alpha}$ at fixed $q=0$ (expressed as $V_{\alpha}|_{q=0}$)
corresponding to two enantiomers with different inner dressed states.
The parameters are
$\Delta/2\pi =100 \, \text{MHz}$,
$\Omega_{12}/2\pi =e^{-z^{2}/\sigma^{2}}\,\text{MHz}$,
$\Omega_{23}/2\pi =10 \times e^{-\left(x+x_{0}\right)^{2}/\sigma^{2}}\,\text{MHz}$, and
$\Omega_{13}/2\pi =10 \times e^{-\left(x-x_{0}\right)^{2}/\sigma^{2}}\,\text{MHz}$
with $\sigma=0.54\,\text{m}$ and $x_{0}=0.16\,\text{m}$.}
\label{fig:va}
\end{figure}
\begin{eqnarray}
\Omega_{12} & = & \tilde{\Omega}_{12}e^{-\left(z-z_{12}\right)^{2}/\sigma_{12}^{2}},\nonumber \\
\Omega_{13} & = & \tilde{\Omega}_{13}e^{-\left(x-x_{13}\right)^{2}/\sigma_{13}^{2}},\label{eq:omega absolute}\\
\Omega_{23} & = & \tilde{\Omega}_{23}e^{-\left(x-x_{23}\right)^{2}/\sigma_{23}^{2}}.\nonumber
\end{eqnarray}
Here $\tilde{\Omega}_{jl}$ are assumed to be positive real constants for simplicity.
Considering the limitation of large detuning and weak coupling,
we choose
$\tilde{\Omega}_{12}/2\pi =1\,\text{MHz}$,
$\tilde{\Omega}_{23}/2\pi =10\,\text{MHz}$,
$\tilde{\Omega}_{13}/2\pi =10\,\text{MHz}$, and $\Delta/2\pi =100\,\text{MHz}$.
In Eq.~(\ref{eq:omega absolute}),
$\sigma_{jl}$ represents the beam radius of the electromagnetic field.
We assume that the beam radii of all the electromagnetic waves are equal for simplicity
and choose
$\sigma_{jl}=\sigma=0.54\,\text{m}$.
In addition, we assume the center position $x_{13}=-x_{23}=x_{0}=0.16\,\text{m}$ and $z_{12}=0\,\text{m}$.

Figures~\ref{fig:va}(a) and \ref{fig:va}(b)
show the induced scalar potentials $V_{\alpha}$ at fixed $p=0$ (expressed as $V_{\alpha}|_{p=0}$)
corresponding to two enantiomers with the inner dressed states. It is obvious that
$V_{\alpha}|_{p=0}$ exhibits the slowly varying envelope multiplied by a quickly varying quasi periodic function.
For the same inner dressed state, the envelope of $V_{\alpha}|_{p=0}$ is chirality independent.
In order to discuss the difference in the quasi periodic function of two enantiomers,
Figs.~\ref{fig:va}(a) and ~\ref{fig:va}(b) are enlarged and shown together in Fig.~\ref{fig:va}(c) given in the space interval
$q\in [-0.04,\,0.04]\,\text{m}$, which is of the wavelength scale of electromagnetic fields.
For the same inner dressed state, the quasi-periodic structure of $V_{\alpha}|_{p=0}$ of two enantiomers is chirality dependent and differs by a half-integer period.

Figure~\ref{fig:va}(d) shows the induced scalar potentials $V_{\alpha}$ at fixed $q=0$ (expressed as $V_{\alpha}|_{q=0}$)
corresponding to two enantiomers.
Here $V_{\alpha}|_{q=0}$, which is not affected by the space-dependent part $\phi_{r}$ of the overall phase, presents chirality-dependent features similar to the case in Ref.~\cite{Li2}, where the molecular rotations were not considered.
Comparing Fig.~\ref{fig:va}(c) with Fig.~\ref{fig:va}(d) in a typical wavelength range (about the order of $0.01\,\text{m}$) of electromagnetic fields,
we find that, while $V_{\alpha}|_{q=0}$ in Fig.~\ref{fig:va}(d) is almost unchanged in the $p$ direction, $V_{\alpha}|_{p=0}$ in Fig.~\ref{fig:va}(c) varies quickly in the $q$ direction.
This affects the center-of-mass motion of chiral molecules.

We see from Eq.~(\ref{eq:A and V}) that
the induced vector potentials in the system under consideration
are $\vec{A}_{\alpha}=A_{\alpha,q}(q,p)\hat{q}+A_{\alpha,p}(q,p)\hat{p}$.
We obtain the effective magnetic field $\vec{B}_{\alpha}=\nabla\times\vec{A}_{\alpha}$, which
has only one component $B_{\alpha,y}(q,p)\hat{y}$.
Here $B_{\alpha,y}|_{p=0}$ [i.e., $B_{\alpha,y}(q,p)$ at fixed $p=0$] presents the chirality-dependent quasi periodic features with the chirality-independent envelope,
while $B_{\alpha,y}|_{q=0}$ [i.e., $B_{\alpha,y}(q,p)$ at fixed $q=0$] presents chirality-dependent features very similar to those of the case in Ref.~\cite{Li2}.
The feature of $B_{\alpha,y}$ is similar to that of $V_{\alpha}$, but its effect on the center-of-mass motion of molecules is much smaller than that of $V_{\alpha}$ under the typical parameters considered here. In the following we show the center-of-mass motion of the chiral molecule governed by the above-mentioned induced gauge potentials.

\section{spatial enantioseparation} \label{Optimization}
So far, we have shown how the space-dependent part $\phi_{r}$ of the overall phase affects the induced gauge potentials.
In this section
we explore the center-of-mass motion of chiral molecules. Due to the large molecular mass and weak effective gauge potentials,
the molecular propagation trajectory
is treated classically~\cite{Li2}. Then the center-of-mass motion is governed by~\cite{Li2}
\begin{figure}[htbp]
\centering
\includegraphics[width=0.9\columnwidth]{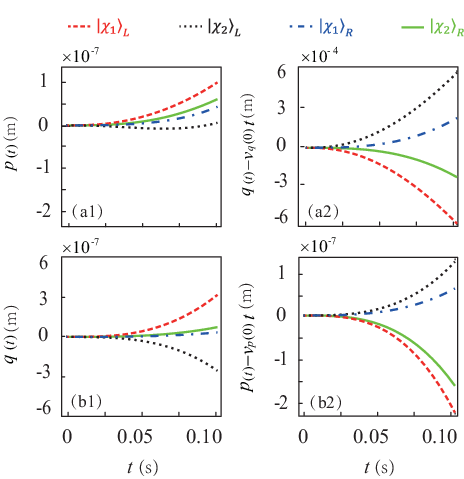}
\caption{
Motion of the chiral molecules:
(a1) and (b1) the displacement perpendicular to the direction of the initial velocity; (a2) and (b2) the displacement increment in the direction of the initial velocity. The molecules are initially placed at $(q(0),p(0))=(0,0)$ with a speed of $|\vec{v}(0)|=0.1\,\text{m/s}$. The initial velocity is in (a) the $\hat{q}$ direction and (b) the $\hat{p}$ direction. The parameters of the electromagnetic fields are the same as those in Fig.~\ref{fig:va}.
\label{fig:Short}}
\end{figure}
\begin{eqnarray}
\dot{q} & = & v_{q}-\frac{A_{q}}{m},\nonumber \\
\dot{p} & = & v_{p}-\frac{A_{p}}{m}, \\
\dot{v_{q}} & = &
\frac{1}{m}\left[\left(v_{q}-\frac{A_{q}}{m}\right)\partial_{q}A_{q}+\left(v_{p}-\frac{A_{p}}{m}\right)\partial_{q}A_{p}-\partial_{q}V\right],\nonumber \\
\dot{v_{p}} & = &
\frac{1}{m}\left[\left(v_{q}-\frac{A_{q}}{m}\right)\partial_{p}A_{q}+\left(v_{p}-\frac{A_{p}}{m}\right)\partial_{p}A_{p}-\partial_{p}V\right].\nonumber
\label{eq:xz2}
\end{eqnarray}
Here $v_{q}$ and $v_{p}$ represent the components of the molecular velocity $v$ in the $\hat{q}$ and $\hat{p}$ directions, respectively.
Since the equations of the motion for different inner dressed states are independent, we omit the subscript $\alpha$ for $q_{\alpha}$, $p_{\alpha}$, $v_{\alpha,q}$, $v_{\alpha,p}$, $A_{\alpha,q}$, $A_{\alpha,p}$, and $V_{\alpha}$.
The molecules are initially placed at $(q(0),p(0))=(0,0)$ with
the initial speed $|\vec{v}(0)|=0.1\,\text{m/s}$.

In Fig.~\ref{fig:Short} we show the motion of two enantiomers in the time interval $t\in[0,\,0.1]\,\text{s}$, where the chiral molecules
move in the distance of a typical wavelength scale.
We are interested in two cases with the initial velocity along $\hat{q}$ direction [shown in Figs.~\ref{fig:Short}(a1) and \ref{fig:Short}(a2)] and in the $\hat{p}$ direction [shown in Figs.~\ref{fig:Short}(b1) and \ref{fig:Short}(b2)], respectively.
Figures~\ref{fig:Short}(a1) and \ref{fig:Short}(b1) show the change of the displacement perpendicular to the initial velocity direction over time in these two cases, respectively. We find that these displacements in Figs.~\ref{fig:Short}(a1) and \ref{fig:Short}(b1) are
much smaller than the corresponding displacement in the initial velocity direction (about $0.01\,\text{m}$).
Then the movement is regarded as one dimensional in the direction of the initial velocity.

The one-dimensional movement in the direction of the initial velocity reflects some characteristic of the induced scalar potentials in this direction.
Thus, we show the change of the displacement increment
in the direction of the initial velocity over time in these two cases in Fig.~\ref{fig:Short}(a2) and \ref{fig:Short}(b2), respectively.
For the same enantiomer with the same inner dressed state,
comparing the displacement increment in the initial velocity direction in Figs.~\ref{fig:Short}(a2) and \ref{fig:Short}(b2), the former [i.e., $q(t)-v_{q}(0)t$] is thousands of times longer than the latter [i.e., $p(t)-v_{p}(0)t$].
Comparing the separation distance between two enantiomers with the same inner dressed state in Fig.~\ref{fig:Short}(a2) with the one in Fig.~\ref{fig:Short}(b2), the former is also thousands of times longer than the latter.
In this sense, we conclude that, when the chiral molecules move the distance of a typical wavelength scale, the separation distance between two enantiomers corresponding to the initial velocity along the $\hat{q}$
axis (affected by $\phi_{r}$) is about $1000$ times longer than that corresponding to the initial velocity along the $\hat{p}$ axis (not affected by $\phi_{r}$).

\begin{figure}[htbp]
\centering
\includegraphics[width=0.9\columnwidth]{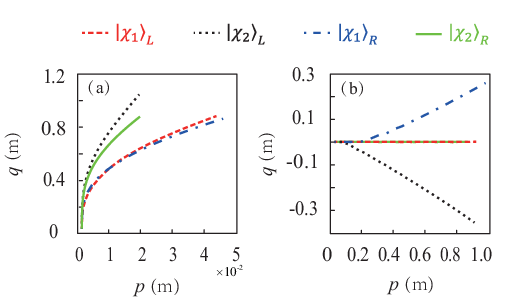}
\caption{
Propagation trajectories of the chiral molecules in the time interval $t\in[0,\,10]\,\text{s}$. The molecules are initially placed at $(q(0),p(0))=(0,0)$ with the speed of $|\vec{v}(0)|=0.1\,\text{m/s}$. The initial velocity is in (a) $\hat{q}$ direction and (b) the $\hat{p}$ direction. The parameters of the electromagnetic fields are the same as those in Fig.~\ref{fig:va}.
\label{fig:Long}}
\end{figure}

The propagation trajectory of chiral molecules in the time
interval $t\in[0,\,10]\,\text{s}$ is shown in Fig.~\ref{fig:Long}, where the
chiral molecules move a distance much longer than a
typical wavelength. The initial velocity is in the $\hat{q}$ direction for
Fig.~\ref{fig:Long}(a) and a $\hat{p}$ direction for Fig.~\ref{fig:Long}(b).
When we let the chiral molecules move in the distance much
longer than a typical wavelength,
the separation distance
between two enantiomers corresponding to the initial velocity along the $\hat{q}$ axis (affected by $\phi_{r}$) is comparable to that corresponding to the initial velocity along the $\hat{p}$ axis (not affected by $\phi_{r}$).

\section{Conclusion} \label{Conclusion}
We have discussed the spatial enantioseparation of gaseous chiral molecules taking into consideration molecular rotations.
Comparing with the original works~\cite{Li2,ShapiroLiX} without consideration of molecular rotations, the overall phase of the coupling strengths has the space-dependent feature due to the specific requirements of the polarization directions of the three electromagnetic fields~\cite{Ye,Koch}.
This space-dependent feature provides an additional resource to generate the chirality-dependent induced gauge potentials. Specifically, the induced gauge potentials vary intensely in the length scale of the typical wavelength of the applied electromagnetic fields. Taking the 1,2-propanediol as an example and using the typical parameters, we have shown that this property of induced gauge potentials provides some advantages for spatial enantioseparation, especially when the chiral molecules move the distance of a typical wavelength scale.

We would like to point out that here we have focused only on
the case of large detuning
in the discussion for simplicity. In fact, our results of spatial enantioseparation can be extended to general cases without the condition of large detuning~\cite{ShapiroLiX}.
In addition, we have only considered that the molecular ensemble is initially located at the same point.
In realistic cases, the molecular ensemble has an initial position distribution in the initial position.  Consideration of the specific limit to the molecular ensemble is left for future work.

\section{Acknowledgement}
This work was supported by the National Key R\&D Program of China (Grant No. 2016YFA0301200), the Science Challenge Project (Grant No. TZ2018003), and the Natural Science Foundation of China (Grants No. 12074030, No. 11774024, No. 11947206, No. 12088101, No. U1930402, and No. U1930403).


\begin{thebibliography}{10}
\bibitem{dk}
Here $\delta\vec{k} \equiv \vec{k}_{13}-\vec{k}_{23}-\vec{k}_{12}$ with $\vec{k}_{jl}$ the wave vectors of the applied electromagnetic fields.
The phase matching is satisfied if $\delta\vec{k} = 0$.
\end{thebibliography}
\end{document}